\documentclass[prl,aps,twocolumn,showpacs]{revtex4} 
\usepackage{graphicx}
\begin{document}

\def\ii{\'\i}
\def \st {\widetilde \sigma}
\def \sd {\sigma ^{\dag}}
\def \pt {\widetilde\pi}
\def \pd {\pi ^{\dag}}
\def \fn {$\Phi_{NLM}$}
\def \fav {$\Phi_{A^{'}}$}
\def \fx {$\Phi _{x}$}
\def \fa {$\Phi _{A}$}
\def \np{n_{\pi}}
\def \f7 {$f_{7/2}$}
\def \si {$\sigma$}
\def \oa {$^{16}O + \alpha$ \ }
\def \ca {$^{12}C + \alpha$ \ }
\def \cam {$^{12}C + \alpha$}
\def \co {$^{12}C + \ ^{16}O$ \ }
\def \com {$^{12}C + \ ^{16}O$}
\def \ma {$^{24}Mg + \alpha$}
\def \mam {$^{24}Mg + \alpha$ \ }
\def \cc {$^{12}C + \ ^{12}C$ \ }
\def \ust {$U^{ST}(4)$ \ }
\def \ustc {$U^{ST}_C(4)$ \ }

\pagebreak
\title{Multichannel dynamical symmetry and cluster-coexistence}
\author{J.~Cseh}
\affiliation{Institute for Nuclear Research, Hungarian Academy of Sciences, Debrecen, Pf. 51, Hungary-4001}
i\author{K.~Kat\=o}
\affiliation{Nuclear Data Centre, Faculty of Science, Hokkaido University, Sapporo, 060-0810, Japan}
%\author{A.~Gregorieva}
%\affiliation{Department of Physics, University of York, Heslington, York YO10 5DD, United Kingdom}
\date{\today}

\begin{abstract}
{A composite symmetry of the nuclear structure, called multichannel dynamical symmetry
is established. It can describe different cluster configurations (defined by different reaction channels)
in a unified framework, thus it has a considerable predictive power. 
The two-channel case is presented in detail, and its
conceptual similarity to the dynamical supersymmetry is discussed.}
\end{abstract}

\maketitle

%\pacs{21.10.Re, 21.60.Cs, 21.60.Fw, 21.60.Gx, 27.30.+t}    

%\section{Introduction}
Symmetry-considerations can simplify the solution of complex problems to a 
large extent. In particular, they turned out to be very useful in the
study of nuclear structure, too. Symmetries of a special kind,
which we call here composite symmetries, give a unified description
of systems of coupled degrees of freedom. Notable examples in
nuclear physics are the pioneering supermultiplet theory by Wigner
\cite{wig37},
accounting for proton and neutron degrees of freedom,
and more recently the dynamical supersymmetries (SUSY), describing
the collective motion and the single nucleon degrees of freedom 
\cite{ivi,fji,misusy}.

The coexistence of different cluster configurations in a nucleus
is an interesting example of a system of coupled degrees of freedom.
Remarkable examples are the 
$^{12}$C+$^{12}$C and $^{20}$Ne+$^{4}$He configurations in $^{24}$Mg, or the
$^{12}$C+$^{16}$O and $^{24}$Mg+$^{4}$He configurations in $^{28}$Si.
In \cite{musy}
the multichannel dynamical symmetry (MUSY) was introduced for the 
description of this phenomenon.
Here the channel refers to the reaction channel, which defines a
binary cluster configuration. 
The idea was invented on the phenomenological level,
based on general physical arguments, which resulted in  relations  
for the energy-eigenvalues 
%(and other observables)
\cite{musy}.
The mathematical background, and the exact physical nature
of this new symmetry has, however,  not been revealed so far. 
Here we present the scenario, how one can establish the algebraic
structure of the MUSY  in general, and give a detailed derivation
for the two-channel dynamical symmetry.

In order to illustrate the main features of this new symmetry it seems to be proper
to recall  some basic vocabulary on symmetries.
A continuous symmetry is an exact one if the Hamiltonian commutes with 
the generators of its Lie-group.
A dynamical symmetry is said to hold if the Hamiltonian 
%of the system 
can be expressed
in terms of the invariant operators of a
chain of nested subgroups (see e.g. 
\cite{ell,ia,rowe}):
%\begin{equation}
%$
G $\supset$ G$^{\prime}$ $\supset$ G$^{\prime \prime}$ $\supset$ ... \  G$^{f}$. 
%$
%\label{eq:group}
%\end{equation}
(Sometimes this symmetry is called dynamically broken one, because only 
G$^{f}$ is an exact symmetry.)
In such a case the eigenvalue problem of the energy  has an
analytical solution, the labels of irreducible representations
are  good quantum numbers.
%, but the degeneracy corresponding
%to an exact $G$ dynamic symmetry splits up. 
%(The Hamiltonian is not scalar with respect to the symmetry-groups, but its eigenvecors are
%still symmetric i.e. they transform according to some irreducible representations (irreps)
%\cite{varna}.)

Let us consider a system of two components ($1$ and $2$), 
each of them described by an algebraic model, and having (at least one)
dynamical symmetry:
%\begin{equation}
%$
G$_i$ $\supset$ G$^{\prime}_i$ $\supset$ G$^{\prime \prime}_i$ $\supset$ ... ;
\ \ i = 1,2 .
%$ 
%\label{eq:gi}
%\end{equation}
If the particle number of the two components are conserved separately,
then the algebraic model with group structure
%\begin{equation}
%$
G$_1$ $\otimes$ G$_2$
%$
%\label{eq:prod}
%\end{equation}
usually proves to be a successful approach.
The subgroups of 
%$
G$_1$ $\otimes$ G$_2$
%$
define the relevant dynamical symmetries of the system.

Deeper symmetries, with different nature arise from the
embedding of the direct product group
%$G_1 \otimes G_2$
into a larger group:
%\begin{equation}
%$
G$_0$ $\supset$ G$_1$ $\otimes$ G$_2$ .
%$
%\label{eq:embed}
%\end{equation}
Some generators ofG$_0$ transform particles of type $1$ into 
particles of type $2$, or vice versa.
%, thus giving rise to a much wider symmetry.
%, having stronger restrictions on the physical operators,
%and providing larger multiplet structure.
In this paper we refer to a composite symmetry in this sense.
In the supermultiplet scheme of Wigner e.g. the protons and neutrons are not
conserved separately
%, transformations are allowed, which take
%them into each other. 
\cite{wig37,fji}.
In the SUSY models collective phonons and nucleons are  transformed
into each other, due to the embedding into graded Lie-algebras
(U(6/m) $\supset$ U(6) $\otimes$ U(m) in the quadrupole
\cite{ivi,fji}, and
U(4/m) $\supset$ U(4) $\otimes$ U(m) in the dipole, i.e. cluster models
\cite{misusy}.)

The multichannel dynamical symmetry is formulated in the framework of the
semimicroscopic algebraic cluster model (SACM) 
\cite{sacm}, in which
the clusterization of atomic nuclei is described in a fully algebraic way. 
%This means that not only the basis states have well-defined symmetry 
%properties, but the interactions (and other physical  quantities) as well.
The model-space is constructed microscopically,
thus one can take into account  that the antisymmetrization may wash out the difference
between different cluster configurations.

The logic of the MUSY, illustrated by the two-channel case, is as follows.
i) First we show that the two binary clusterizations are related to each other
by an underlying ternary configuration, which has two different sets
of the relevant Jacobi coordinates. ii) The transformation between the two sets
is well-established, and has definite algebraic structure
(obtained by the extension of the results of
\cite{bamo,kn,gal,hor77}). 
iii) A chain of nested subgroups enables us to define
a dynamical symmetry of the ternary configuration.
% in the usual sense.
%(as discussed above). 
iv) The two-channel dynamical symmetry
of the different binary confgurations is obtained from the dynamical
symmetry of the ternary configuration by projection.
% i.e. by applying appropreate contstraints. 
This rigorous derivation results in the same energy-functional,
which was obtained in a phenomenological way in 
\cite{musy}.

In what follows first we recall the basic features of the SACM
%semimicroscopic algebraic cluster model, 
and the empirical introduction of the MUSY,
then we go step-by-step in its rigorous derivation. 
The example we consider is the 
$^{16}$O+$^{8}$Be and $^{20}$Ne+$^{4}$He configurations of $^{24}$Mg,
in which case each cluster of the underlying ternary configuration
($^{16}$O+$^{4}$He+$^{4}$He)
 is SU(3) scalar.
 Finally we 
compare  some characteristic features of this new symmetry  to
those of the  dynamical supersymmetry.

%\section{The semimicroscopic algebraic cluster model}

%\subsection{Basic concepts}
{\it The semimicroscopic algebraic cluster model}
\cite{sacm} 
describes the internal structure of the clusters  by the Elliott model
\cite{ell},
therefore, its wave function has a
U$^{ST}_C$(4) $\otimes$ U$_C$(3)
symmetry, where $C$ indicates cluster.
% (internal structure).
The relative motion of the clusters is accounted for by the modified vibron model
\cite{vibron}.
This is an algebraic model of the rotation and vibration of a two-body system  in 
the three dimensional space, which has a U$_R$(4) group structure
($R$ stands for relative motion).
The spin and isospin degrees of freedom are essential
in constructing  the model space. However, if
one is  interested only in a single  supermultiplet (U$^{ST}_{C}$(4))
symmetry, which is typical in cluster problems, then 
from the viewpoint of the operators
the relevant group structure
%, from the viewpoint of the operators, 
simplifies to that of the space part:
%In particular  the
%$U(3)$ (strong) coupled (space part) basis is defined by the
%group chain:
\begin{widetext}
\begin{eqnarray}
 U_C(3) \ \ \otimes \ \ U_R(4) \ \ \ \supset U_C(3)
 \otimes U_R(3) \ \
 \supset
\  \ \ U(3) \ \supset SU(3) \supset SO(3) \supset SO(2)
 \nonumber \\
 \vert  [n^C_1,n^C_2,n^C_3] \ ,  [N_R,0,0,0]  ,\ \ \ \ \ \ \ \ \ \
 [n_{R},0,0,] \  ,  \  
%\rho ,
 [n_1,n_2,n_3]  ,
 (\lambda , \mu) , K \ \ ,\ L \ \ \ \ ,\ \ \  M \ \rangle .
 \label{eq:omusy}
\end{eqnarray}
\end{widetext}
%and the physical operators are constructed from their generators.
%Here the quantum numbers characterizing the basis states are also indicated.
The coupled U(3) group is generated by the:
%\begin{equation}
$
n=n_{C} +  n_{R} , \ 
Q = Q_C + Q_R , \  
L = L_C + L_R  ,
$
%\label{eq:unitgencl}
%\end{equation}
operators, 
where $n$ is the particle (quantum) number operator, $Q$ is the 
quadrupole momentum, and $L$ is the angular momentum.

This strong coupled  U(3)  basis is especially useful
for treating the exclusion principle
\cite{sacm}, 
since the U(3) generators commute
with those of the permutation group
%, therefore, all the basis sates of an  irrep are either Pauli-allowed, or forbidden
\cite{hos62}.
The exclusion of the Pauli-forbidden states 
is a modification
\cite{sacm} 
with respect to the original
vibron model, as it is applied e.g. in molecular physics
\cite{vim}.  

%\subsection{Phenomenological introduction of MUSY}
The argumentation of
\cite{musy}
introducing the MUSY on the level of eigenvalues is as follows. 
Let us consider two different binary clusterizations of a nucleus.
For the sake of simplicity let each cluster be spin-isospin scalar,
and one cluster in both configurations be SU(3) scalar (having closed shell
structure). 
An example 
is the $^{20}$Ne +  $^4$He and  $^{16}$O + $^{8}$Be configurations of the
$^{24}$Mg nucleus.
What is the relation between the energy spectra of the two configurations,
if the dynamical symmetry
(1)
%( \ref{eq:omusy})
holds
%(to a reasonable approximation) 
for both of them?
The U(3) basis states of the different clusterizations are not orthogonal 
to each other, they may have considerable overlap as 
a consequence of the antisymmetry of the wavefunction.
In some cases the two wavefunctions can even be
%(apart from a normalization factor)
identical with each other.
Therefore, it is natural to {\sl require} that their energies be the same, 
too.
This requirement 
establishes a relation between the energy-eigenvalues of the
two Hamiltonians.
In 
\cite{musy}
the energy functional
\begin{equation}
E = \epsilon + \gamma n_{R} + \beta L(L+1) + \theta n_{R}
L(L+1) + F(\lambda , \mu ,L)
\label{musen}
\end{equation}
was applied for the spectra of two different ($c$ and $d$) binary clusterizations
with the constraints of
%\begin{eqnarray}
$
 \gamma _c =  \gamma _d =  \gamma ,
 \  \epsilon _c =  \epsilon _d + \gamma  n_{{0}},
 \ \theta _c = \theta _d = \theta ,
%\nonumber \\
 \ \beta _c = \beta _d + \theta n_{{0}},
\nonumber \\
 \ F_c(\lambda , \mu ,L) =  F_d(\lambda , \mu ,L), 
\ _cn_{{R}} = {_dn}_{{R}} + {n}_{{0}} \ \ \ \  \ \ \ \ \ \ \ \ \ \ \ 
\label{eq:nfug}
%\end{eqnarray}
$
where 
${n}_{{0}}$
is the difference of the relative motion quantum number in describing
the same SU(3) state.
%:$_cn_{{R}} = {_dn}_{{R}} + {n}_{{0}}$.

%\section{Microscopical foundation of  MUSY}
%Here we give a detailed derivation of the two-channel dynamical symmetry,
%but  the scenario is similar for other cases, too.  

%\subsection{Underlying ternary configuration}
{\it The microscopical foundation of the MUSY}
starts with the introduction of 
an underlying multicluster configuration. For the two-channel case
it is a ternary configuration.
We consider two different binary clusterizations: 
%\begin{equation}
$
c: C_1 + C_2 , \  d: D_1 + D_2 ,
$
and suppose that the relation of the mass-numbers  are:
%\begin{equation}
$
A_{C_1} \ge A_{C_2} , \ 
A_{D_1} \ge A_{D_2} , \ 
A_{D_1} \ge A_{C_1} , \
A_{C_2} \ge A_{D_2} .
$
%\end{equation}
%what can be done without any loss of generality.
(In the example of the 
$^{24}$Mg nucleus,
% mentioned before,
$C_1$:\ $^{16}$O, $C_2$:\ $^{8}$Be,
$D_1$:\ $^{20}$Ne, $D_2$:\ $^{4}$He.)
Let us consider the following ternary fragmentation
%\begin{equation}
$
(C_1)+(CD)+(D_2), \ 
(CD) = (C_2 - D_2) = (D_1 - C_1) .
$
%\end{equation}
(In the example: $^{16}$O+$^{4}$He+$^{4}$He.)
Another ternary clusterization is given by:
%\begin{equation}
$
(C_2)+(DC)+(D_2), \ 
(DC) = (C_1 - D_2) = (D_1 - C_2) .
$
%\end{equation}
(In the example: $^{8}$Be+$^{12}$C+$^{4}$He.)

We suppose that each fragment  has a definite U(3)
symmetry (intrinsic state), and consider
the ternary fragmentation, which is simpler in the sense
that it contains more SU(3)-scalar clusters,  e.g.
$
(C_1)+(CD)+(D_2).
$
Then the two sets of Jacobi coordinates, defining the binary configurations
are:
$
{\bf t}_c = {\bf r}_{D_2} - {\bf r}_{CD}, \
 {\bf s}_c = {\bf r}_{C_1} -(M_{D_2}{\bf r}_{D_2} + M_{CD}{\bf r}_{CD})/(M_{D_2} + M_{CD}), 
$
and
$
{\bf t}_d = {\bf r}_{C_1} - {\bf r}_{CD},   \
{\bf s}_d = 
{\bf r}_{D_2} -
(M_{C_1}{\bf r}_{C_1} + M_{CD}{\bf r}_{CD}) / (M_{C_1} + M_{CD}), 
$
where $M$ is the mass and ${\bf r}$ is the space vector
of the corresponding cluster.
Obviously, the clusterization
$C_1 + C_2$, 
corresponds to the coordinate-set $c$
with some restriction on ${\bf t}_c$,
while clusterization $D_1 + D_2$
corresponds to the coordinate-set $d$
with some restriction on ${\bf t}_d$.

The transformation from the clusterization
$C_1 + C_2$ to that of $D_1 + D_2$
requires a transformation between 
the two sets of Jacobi coordinates:
${\bf s}_c, {\bf t}_c$ and
${\bf s}_d, {\bf t}_d$.

%A similar set of coordinates and their transformation can be found 
%for the other ternary clusterization as well.

%\subsection{Dynamical symmetry of the ternary configuration}
For the description of the two independent relative motions along the 
$
{\bf s}
$
and
$
{\bf t}
$
vectors we introduce two sets of oscillator quanta (i.e. $l=1$ bosons). 
The corresponding  creation operators 
%\begin{equation}
$
_{i}\pd _{\mu}, \ ( \mu = -1,0,1,  \ i=1,2) ,
$
%\end{equation}
and annihilation operators 
$
_{i}\pi _{\mu} ,
$
satisfy the commutation relations
$ [_{i}\pi _{\mu} , { _{j}\pd _{\nu}} ] =
\delta_{ij}\delta_{\mu \nu}$. 
Here 1 and 2 refer to the sets of quanta along the 
$
{\bf s}_k
$
and
$
{\bf t}_k
$
($k=c,d$) coordinates.
Furthermore, an ($l=0$) scalar $\sigma$ boson is introduced
in order to be able to generate the spectrum
\cite{bil}.
%, similarly to the case of the simple vibron model.
%The spherical tensor operators are obtained as in Eq.
%(\ref{boper}).
The particle number conserving bilinear products, 
coupled to good spherical tensors, are:
\begin{equation}
 \begin{array}{l}
 \displaystyle{
_{ii}B^{(l)}_m (1,1) = [ _i{\pi}^{\dag} \times  \widetilde {_i{\pi}}
]^{(l)}_m , \ \ \ \ i=1,2 \ , } \\ 
 \displaystyle{
_{12}B^{(l)}_m (1,1) = [ _{1}{\pi}^{\dag} \times  \widetilde {_{2}{\pi}}
]^{(l)}_m ,  \
%} \\ 
% \displaystyle{
_{21}B^{(l)}_m (1,1) = [ _{2}{\pi}^{\dag} \times  \widetilde {_{1}{\pi}}
]^{(l)}_m ,  } \\
 \displaystyle{
_{0i}B^{(l)}_m (0,1) = [{\sigma}^{\dag} \times  \widetilde {_{i}{\pi}}
]^{(l)}_m \ \ , \ \ 
%($CD$) 
%} \\
% \displaystyle{
_{i0}B^{(l)}_m (1,0) = [_i{\pi}^{\dag} \times  \widetilde {{\sigma}}
]^{(l)}_m   } , \\
 \displaystyle{
_{00}B^{(0)}_0 (0,0) = [{\sigma}^{\dag} \times  \widetilde {{\sigma}} 
]^{(0)}_0   } . 
\label{eq:gener}
\end{array}
\end{equation}
Here 
$\widetilde {_{i}{\pi}_{\mu}} = (-1)^{(-\mu)} {\pi}_{\mu} , \ 
\widetilde {{\sigma}} = \sigma $,
and the $[\  ]$ brackets indicate angular momentum coupling. 
The $2 \times 9$ operators of the first line generate 
two U(3) groups.
Together with the other $2 \times 9$ operators of 
the second line, they generate an U(6) group.
With the $2 \times 6 +1$ operators containing (also) $\sigma$
bosons the U(7) group is obtained.

This group has a rich structure of subgroups,
therefore, several dynamical symmetries can be constructed.
The important one (for the present purpose) 
starts with the 
U(7) $\supset$ U(6) group-chain, and contains a
unified U(3) group and a pseudo-spin group
U$_p$(2)
\cite{gal,kn}.
U(3) is 
generated by the operators:
%\begin{equation}
$
B^{(l)}_m (1,1) = 
{_{11}B^{(l)}_m (1,1)} + 
{_{22}B^{(l)}_m (1,1)} , 
\label{u3gen}
%\end{equation}
$
while the U$_p$(2)  generators are:
\begin{equation}
 \begin{array}{l}
 \displaystyle
{
S^{p}_{-} =
\sum_m \ {_2\pd _m} \ {_1\pt _m} \ , \
% =
%{\sqrt {3}} \ {_{21}B^{(0)}_0 (1,1)} ,
%} \\
% \displaystyle
%{
S^{p}_{+} =
\sum_m \ {_1\pd _m} \ {_2\pt _m} 
%=
%{\sqrt {3}} \ {_{12}B^{(0)}_0 (1,1)} ,
} \\
 \displaystyle
{
S^{p}_{0} = {1 \over 2}
\sum_m \ ( {_1\pd _m} \ {_1\pt _m} -
           {_2\pd _m} \ {_2\pt _m} ) 
%\ = \
%{1/2} \ ( {_1n_{\pi}} - {_2n_{\pi}} ) ,
%} \\
 %\displaystyle
%{
%\ \ \ \ \ \ \ \ \
%= {{\sqrt {3}}/2} \ \lbrace {_{1}B^{(0)}_0 (1,1)} -
%                               {_{2}B^{(0)}_0 (1,1)} \rbrace ,
} \\
 \displaystyle
{
N_6 = 
\sum_m \ ( {_1\pd _m} \ {_1\pt _m} +
           {_2\pd _m} \ {_2\pt _m} ) 
 = \ n_1 + n_2 .
%= \ {_1n_{\pi}} + {_2n_{\pi}} 
%} \\
% \displaystyle
%{
%\ \ \ \ \ \ \ \ 
%= {{\sqrt {3}} } \ \lbrace {_{1}B^{(0)}_0 (1,1)} +
%                                {_{2}B^{(0)}_0 (1,1)} \rbrace .
} 
\end{array}
\end{equation}
Here
$\sum_m \ {_i\pd _m} \ {_j\pt _m} =
{\sqrt {3}} \ {_{ij}B^{(0)}_0 (1,1)} .$
The pseudo spin operators, first introduced by Bargman and Moshinsky 
\cite{bamo}
in a different context, act in the cluster-index space
\cite{hor77}, 
and an exact  
U$_p$(2) symmetry means complete invariance with respect to all
the transformations in the particle index space,
including the finite rotation 
from the set of Jacobi coordinates $c$
to that of $d$.

The dynamical symmetry of the ternary configuration is described by the group-chain:
\begin{widetext}
\begin{eqnarray}
 U(7) \supset \ U(1) \otimes U(6) \supset \ U(6) \supset 
 \lbrace
 U(3) \supset SU(3) \supset SO(3)  \supset SO(2)
 \rbrace 
 \otimes 
 \lbrace
 U_p(2) \supset SU_p(2) \supset SO_p(2) 
 \rbrace \\
 \nonumber 
 \vert  N_7 \ \ , \ \ \ \ n_{\sigma} \ \ \ \ \ \ \ \ \ \ \ \ \ \ \ \ 
 N_6 \ , \ \ \ 
 [h_1 ,h_2 , 0] ,  
 (\lambda , \mu) ,  K, \ \  L \ \ \ , \ \ \ \ M , \ \ \ \ \ \ \ \
 [h_1 ,h_2] , \ \ \ \ \ \ j_p \ \ \ \ \ , \ \ \ \  \ m_p 
\rangle . 
 \label{musych}
\end{eqnarray}
\end{widetext}
The relation of the representation labels are:
%\begin{equation}
 %\begin{array}{l}
 %\displaystyle{
$ 
N_6 = N_7,  N_7 - 1 , N_7 - 2 , ... , N_0; \
 %}  \\
% \displaystyle{
 h_1 \ge h_2 , \ h_1 + h_2 = N_6 = n_1 + n_2; \ 
% }  \\
 %\displaystyle{
 j_p = {1 \over 2} (h_1 - h_2); \ 
  m_p = j_p , j_p - 1 , ... - j_p;  \
 %\displaystyle{
 m_p = {1 \over 2} (n_1  - n_2);   \
 %}  \\
 %\displaystyle{
 [h_1 , h_2] = [n_1  + n_2, 0] , [n_1 + n_2  - 1, 1] , ... ,  \
 %\displaystyle{
 %\ \ \ \ \ \ \ \ \ \ \ \ \ \  
 [max \{n_1, n_2\} , min \{n_1 , n_2\}] ; \
 %}  \\
 %\displaystyle{
 \lambda = h_1 - h_2, \  \mu = h_2 ; \
 %}  \\
 %\displaystyle{
 K = min \{ \lambda ,\mu \} , min \{ \lambda ,\mu \} -2 , ... , 
 1 \ or \ 0 ; \
% }  \\
 %\displaystyle{
 if \  K = 0: L = max \{ \lambda ,\mu \} , 
 max \{ \lambda , \mu \} - 2 , ... , 1 \ or \ 0 , \
% } \\
 %\displaystyle{
 if \ K \not= 0: L = K , K+1 , ... , K+max \{ \lambda ,\mu \}.
$
$N_0$ is the lowest Pauli-allowed value.
% }
%\end{array}
%\end{equation}

Omitting the redundant labels as well as  $N_7$, which is 
taken to be  a constant
for a system, the basis states can be denoted in different ways:
%\begin{eqnarray}
%\begin{equation}
$
\vert  N_6,  [h_1 ,h_2],  m_p , K  ,L  \rangle , \ 
 \vert  N_6, m_p,  (\lambda , \mu),  K, L  \rangle
$,
%\nonumber  \\
$
\vert  n_1, n_2,  (\lambda , \mu),  K, L  \rangle.
% \ \ \ \ \ \ \ \ \ \ \ \ \ \ \ \ \ 
$
%\end{eqnarray}
%\end{equation}
This latter notation coincides with that of
\cite{hor77}.

Note that the pseudo-spin operators are U(3) $\supset$ SU(3) $\supset$ SO(3) scalars,
while the generators  of U(3) are scalars with respect to the pseudo spin.
Therefore, these latter ones are invariant
under the  transformations in the cluster index space.
%, e.g. from the coordinate system 
%$c$ into $d$.
% (or the other way around).

%This group-chain separates the operators of the space part and the particle (cluster)
%index part. As for the different sets of Jacobi coordiates 
The transformation of the building blocks (creation and annihilation) operators,
between the different sets of Jacobi coordinates
can be found in 
%e.g.
\cite{hor77}. They determine also the transformation of other physical
quantities. 
The  basis states of one set of coordinates can be expressed
as a finite combination of the other basis,
with the Talmi-Moshinsky coefficients. 
%As shown in \cite{hor77}, it simplifies considerably when an SU(3) basis is applied.
Therefore, the expectation values and overlaps expressed in terms of
coordinates $c$  or $d$ can be transformed into each other. We leave these
general investigations for a separate study; here we concentrate on the
consequences of the dynamical symmetry.

A simple, yet realistic
%\cite{sacmappl}
 Hamiltonian can be written as:
\begin{equation}
%$
H = \epsilon + \gamma N_6 
%+ \delta_R n_{\pi}^2 + 
+ \delta C^{(2)}_{SU(3)} + \beta L^2 + \theta N_6 L^2 ,
%$
\label{hamt}
\end{equation}
which is diagonal in the basis 
%(\ref{musych}).
above, describing a dynamical symmetry of the underlying ternary configuration.
%in the usual sense, as discussed in the introduction.
%\begin{eqnarray}
%\begin{equation}
%E = \epsilon + \gamma n_{\pi} 
%+ \delta (\lambda^2 +\mu^2 +\lambda \mu + 3 \lambda + 3 \mu) \\
%\nonumber
%+ \beta L(L+1)
%+ \theta n_{\pi} L(L+1).
%\label{hamt2}
%\end{equation}
%\end{eqnarray}
Of course, more complicated functional forms of the invariant operators of
U(3) and its subgoups  can
be applied, too, still having the dynamical symmetry.
These Hamiltonians have an exact pseudo-spin symmetry.
 
The same procedure, i.e. separating the symmetries of the particle index pseudo-space
and the coordinate-space can be carried out 
%not only for a ternary, but 
for any multicluster
configuration, too, including the limiting case of the $A$-clusters (where $A$ is the
mass number of the nucleus). Therefore, the prescription for the construction
of symmetric Hamiltonians  is applicable
also for multi-cluster configurations, up to the shell model limit.

%Therefore, they are shared 
%not only by different binary cluster configurations, 
%but by the shell model (monocluster) description as well.

%\subsection{Dynamical symmetries of the binary channels}
The dynamical symmetries of the binary configurations are obtained as projections
from the dynamical symmetry of the underlying ternary configuration.
The clusterization $c$ is specified by a constraint on quantum
number
%of the relative motion along 
$n_{t_c}$
(or in the notation of the group-chain (7)
%(\ref{musych}) 
$m_p \equiv m_c$),
%$\bf t_c$, 
while that of 
configuration $d$ fixes $n_{t_d}$ ($m_p \equiv m_d$).
(In the example: $^{16}$O+$^{8}$Be: $n_{t_c}=4$, $m_c=4$;
 $^{20}$Ne+$^{4}$He: $n_{t_d}=8$, $m_d=0$.)
These quantities:
$n_{t_c}$ and  $n_{t_d}$ 
(or $m_c$ and $m_d$) 
are not good quantum numbers at the same time.
The basis state with a definite $n_{t_c}$ is a 
finite linear combination of those of $n_{t_d}$,
and vice versa. 
Therefore, the two-channel MUSY is a partial symmetry 
\cite{pds}
in the sense
that not each of the quantum numbers are valid in the two-channel multiplets.
%we can not establish a multicahnnel multiplet,
%including states of configurations $c$ and $d$.
But it is remarkable that the pseudo-spin-scalar Hamiltonian
%(\ref{hamt1})
is diagonal in both basis (independent of $m_p$), i.e.
a configuration-independent interaction can be constructed.
%by requiring  an exact pseudo-spin symmetry.

The projection to the binary configurations takes place in the following way.
i) The ternary Hamiltonian of Eq. (9) contains a simple form of the
$F(\lambda, \mu, L)$ function: $C^{(2)}_{SU3}$.
Since $F(\lambda, \mu, L)$ corresponds to a pure shell-model contribution
(it contains no relative motion i.e. $n_R$-dependence), other functions
of this kind would behave in the same way (they are invariant
with respect to the transformations from ternary to binary
configurations). 
ii) The relative motion quantum number $n_{s_c}$ or $n_{s_d}$ of the binary
configurations is obtained from the
$N_6 = n_s + n_t$  ternary quantum numbers, by
fixing $n_t$ for the configurations $c$ and $d$ respectively
%as discussed above: 
(in our example  $n_{t_c}=4$,  $n_{t_d}=8$).
Then the ternary Hamiltonians for the Jacobi coordinates $c$ and $d$ are: 
\begin{eqnarray}
H_c = (\epsilon + 4 \gamma) + \gamma n_{s_c} 
+ \delta C^{(2)}_{SU3} + (\beta + 4 \theta) L^2 + \theta n_{s_c} L^2 , \ \ \ 
\nonumber \\
H_d =(\epsilon +8 \gamma) + \gamma n_{s_d}
+ \delta C^{(2)}_{SU3} +(\beta +8 \theta) L^2 + \theta n_{s_d} L^2.  \ \ \
%\nonumber
\label{hamt2}
\end{eqnarray}
Note that the parameters
$\epsilon , \  \gamma  , \  \delta ,  \ \beta , \  \theta$
do not  depend on  the index $c$ or $d$, they belong to the 
pseudo-spin-invariant ternary Hamiltonian.
On the other hand the  parameters of Eq. (2), i.e. the  
energy-formula of the binary configurations are different for clusterization
$c$ and $d$:  
$\epsilon_c, \  \gamma_c  , ...$ and 
$\epsilon_d, \  \gamma_d  , ...$ 
Calculating the eigenvalues of  the projected ternary Hamiltonians of Eq. (7)
according to Eq. (2) one arrives at the relations:
%\begin{eqnarray}
$
 \gamma _c =  \gamma _d =  \gamma ,
 \  \epsilon _c =  \epsilon _d  - 4 \gamma,
 \ \theta _c = \theta _d = \theta ,
%\nonumber \\
 \ \beta _c = \beta _d - 4 \theta,
%\nonumber \\
 \ F_c(\lambda , \mu ,L) =  F_d(\lambda , \mu ,L) , 
%\ \ \ \  \ \ \ \ \ \ \ \ \ \ \
%\ _cn_{{R}} = {_dn}_{{R}} + {n}_{{0}} \ \ \ \  \ \ \ \ \ \ \ \ \ \ \ 
$
%\label{eq:nfug2}
%\end{eqnarray}
i.e. the projections from the underlying ternary dynamical symmetry result
exactly in the relations of the binary dynamical symmetries, which were
obtained from the phenomenological introduction of the MUSY
\cite{musy}.

Similarly to the Hamiltonian, some other operators e.g. those of the
electric quadrupole and magnetic dipole transitions are also diagonal
in both bases.
%, establishing a one-to-one correspondence between their matrix elements.

%\subsection{Coupling to internal structure}
So far we discussed the relative motion part of the cluster configurations.
When there is (at least one) non-SU(3)-scalar cluster in the
ternary configuration,
then $n_C$  $L_C$ and $Q_C$
also contribute to the operators.
% (e.g. in Eq. (\ref{eq:lq})).
%coupling to the internal degrees of freedom gives rise to
%contributions to the physical operators.
%The only difference with respect to Eq. 
%(\ref{eq:lq})
%is that now we have the operators of the relative 
%motion composed of two parts, as in Eqs. 
%(\ref{eq:npipl},\ref{eq:lpl},\ref{eq:qpl}).
Then the corresponding group-chain is:
%\begin{widetext}
%\begin{equation}
%\begin{eqnarray}
%$
U$_C$(3) $\otimes$ $\lbrace$U$_R$(7) $\supset$ U$_R$(1) $\otimes$ U$_R$(6)$\rbrace$
 %\ \ \ \ \ \\
 %\nonumber
$ \supset$  
 $\lbrace$U$_C$(3) $\otimes$ \ U$_R$(3) $\supset$  U(3) $\supset$ SU(3) $\supset$ SO(3) $\supset$ SO(2)$\rbrace$ 
$ \otimes$ 
$ \lbrace$U$_p$(2) $\supset$ SU$_p$(2) $\supset$ SO$_p$(2)$\rbrace$ . 
%\ \ \ \ \ \ \ \ \ \ \ \  \ \ \ \ \ \ \ \ \ \ \  \  \ \ \ \ \ \ \ \ \ \ \ 
%\nonumber
%\end{equation}
 %\end{eqnarray}
%\end{widetext}
%$
Here U$_C$(3) stands for the internal symmetry group of the 
non-scalar cluster, if there is only a single one, 
and it stands for the coupled symmetry of the internal structures
if there are more non-scalar clusters.

%In such a case the discussion on the dynamical and pseudo spin
%symmetries can be applied, too, with the corresponding coupling
%of $U_C(3)$.

%\subsection{Comparison with the supersymmetry}
{\it Comparing the multichannel symetry with the dynamical supersymmetry} 
the following can be said. 
Both the SUSY and the MUSY  are composite symmetries;
they describe a coupled system with two components:
bosonic and fermionic in the SUSY and different cluster-configurations
in the MUSY.  
Both components have  a dynamical symmetry, and some extra transformations 
connect them, by taking particles from one sector to the other.
They are the super transformation in the former, and Talmi-Moshinsky 
transformations in the latter case. 
%Invariance with respect to these
%extra transformations
% leads to  an extra symmetry of the coupled system.
%The full symmetry emergies from the 
%combined with the  simple dynamical
%symmetries leads to the composite symmetries.
%with these extra symmetries.
% For the dynamical
%supersymmetry the multiplets of the two parts are unified into
%a supermultiplet, while for the multichannel symmetry not each
%quantum numbers have well-defined values for the two components
%(one of them  belongs only to a component of the wavefunction).
The SUSY preserves all the quantum numbers, thus provides a supersymmetric
multiplet, while the MUSY is a partial symmetry: one quantum number
is not valid for both channels at the same time. 
Nevertheless, it is
very interesting that the present  mathematical foundation 
of the MUSY shows the way, how the configuration-independent
cluster-cluster interactions can be constructed, which are 
really invariant with respect to the pseudo-spin transformations.
From this viewpoint the MUSY is more strict;
%: its interactions are symmetric, while those of 
in the SUSY models the interactions are usually not invariant with respect
to the super tansformations
(due to the fact that  the first
subgroup of their chain is already a (direct product) Lie-group,
i.e. its Casimir operator is not necessarily invariant with respect to the
super transformations). 
In this respect the MUSY is more similar
to the SUSY models of the particle physics
\cite{jobr}, 
where this kind of
invariance is usual, while in the nuclear SUSY models it is exceptional.

%\section{Summary and conclusions}

%{\it To sum up:} we have investigated the mathematical background and 
%the physical content of the multichannel dynamical symmetry, which
%connects different clusterization within a single nucleus.
%As a particular example the interrelation of two binary 
%cluster-configurations have been considered in detail, but the 
 %procedure can  be applied to other cases, too.

%The simple relation, which was found phenomenologically 
%between the
%energy-eigenvalues (and other observables) of the two different
%clusterizations
%\cite{musy}
 %turns out to be a consequence of a special symmetry
%of the cluster-cluster interactions. This symmetry has a composite
%nature in the  sense that  both cluster-configurations have 
%a dynamical symmetry, and they are combined with a further symmetry,
%which connects them.
%The transformations of this this connecting symmetry act
%in the cluster-index space, in fact 
%they are related to the Talmi-Moshinsky
%transformation of an underlying ternary clusterization, which can be used
%for building up the two binary configurations.
%The symmetry  is  a dynamical symmetry, according to the
%usual definition: the interactions are expressed in terms of the invariant
%operators of a group-chain. 
%The multichannel dynamical symmetry turns out to be the consequence
%of the dynamical symmetry of an underlying multi-cluster configuration.

%The present paper shows that the phenomenological formulas
%applied successfully for the description of experimental data
%are consequences of a deeper symmetry
 %(of the cluster-cluster interactions).

{\it To sum up}: 
we have shown that a  multichannel dynamical symmetry,
which connects different  cluster configurations in a nucleus can be derived from 
the dynamical symmetry of an underlying multicluster configuration.
%The MUSY is very restrictive, i.e.  it requires  strong correlations between the
%different cluster configurations,
%and due to the same reason it seems to have some predictive power.
It establishes a strict correlation 
between the observables of different clusterizations,
% i.e.
%it is a very restrictive symmetry,
%and  has a considerable predictive power.
e.g. the Hamiltonian of one cluster configuration may completely determine
the energy spectrum of another cluster configuration.
%, without any
%free parameter (and without any ambiguity in the model space, due to
%ts microscopic construction).
%As for the experimental observation of cluster spectra of different
%configurations, it is limited by several physical and technical
%circumstances (like penetrability, level density, energy resolution etc).
%Therefore, detailed comparison between the experimental data and model
%calculations can be expected only in the low-lying (ground-state) region,
%while in the highly excited energy domain,
%one can compare mainly the gross features (like e.g. level density
%of a given spin-parity), not so much the states one-by-one.
%After the intuitive introduction of the multichannel symmetry,
%some applications of this kind
%have been carried out.
%In 
%\cite{musy}
%the 
%$^{24}$Mg + $^{4}$He, and
%$^{16}$O + $^{12}$C
%spectra were described with the same Hamiltonian in a wide energy range.
%Furthermore, the density of $\alpha$-particle (scattering or capture) 
%resonances was predicted 
%by this Hamiltonian, 
%without any parameter fitted 
%in the relevant energy region, and the result turned out to be in good
%agreement with the experimental observation.  
%Similar application have done for the
%$^{28}$Si + $^{4}$He, and
%$^{16}$O + $^{16}$O cluster spectra in
%\cite{s32}.
%In light of the present foundation of the multichannel symmetry it turns
%out that this kind of description of the experimental data is provided by
%a symmetry with well-defined mathematical background and physical
%content (of the cluster-cluster interactions).
The  first tests of the MUSY, performed after the phenomenological introduction
\cite{musy,s32}, 
seem to be promising,
but
%, the available applications of the multichannel symmetry
%is still very limited; 
much work remains to be done in order to check
how well this new symmetry is realized in nuclear spectra.

%{\it To sum up}: we have sown that the multichannel dynamical symmetry,
%which was introduced and applied beforehand on the phenomenological level,
%is a projection of a usual dynamical symmetry of an underlying multicluster-configuration.
%Special attention was paid to the two-channel case, which relates two binary clusterizations.
%The method of constructing the configuration-independent inetractions is, however,
%applicable in other cases as well,  including also the shell-model limit.

%\section*{Acknowledgements}
This work was supported by the OTKA (Grant Nos. K72357, K106035), 
and by the JSPS-MTA Bilateral project (No. 119).
One of the authors (J. Cs.) acknowledges the hospitality of the 
%Graduate School of Science of the 
Hokkaido University, where a part of this work 
has been completed.

\end{document}